\documentclass[aps,reprint,prl,superscriptaddress]{revtex4-1}
\usepackage{graphicx}
\usepackage{dcolumn}
\usepackage{bm}
\begin{document}
\title {Experimental signatures of cosmological neutrino condensation }
\vskip 0.3 in
\author{Mofazzal Azam}
\affiliation{Theoretical Physics Division, Bhabha Atomic Research Centre,
Mumbai, India}

\author{Jitesh R. Bhatt}
\affiliation{Theoretical Physics Division, Physical Research Laboratory,
Ahmedabad, India}

\author{Utpal Sarkar}
\noaffiliation

\begin{abstract}
Superfluid condensation of neutrinos of cosmological origin at a low
enough temperature can provide simple and elegant solution to the problems
of neutrino oscillations and the accelerated expansion of the universe.
It would give rise to a late time cosmological constant of
small magnitude
and also generate tiny Majorana
masses for the neutrinos as observed from their flavor oscillations.
We show that carefully prepared beta decay experiments in the laboratory
would carry signatures of such a condensation, and thus, it would be possible
to either establish or rule out neutrino condensation of cosmological
scale in laboratory experiments.

\end{abstract}
\vskip .2 in
\maketitle
\date{}
The flavour oscillations of neutrinos and the accelerated expansion of
the universe are two fundamental problems in particle physics and cosmology.
The neutrino oscillations indicate that neutrinos
are massive and their masses are slightly different for the different flavors.
On the other hand the accelerated expansion of the universe requires an existence of 
some form of dark energy or a cosmological constant.
Due to the proximity of the dark-energy scale with the neutrino masses,
there have been a lot of efforts in finding an unified solution to both
the problems \cite{various}.

Recently, there has been a lot of interests in neutrino condensation
on the cosmological scales \cite{mavan}. It has been argued that the 
background neutrinos
under certain conditions can become superfluid \cite{caldi, utpal}. 
The neutrino condensates can then 
contribute towards the accelerated expansion of the
 universe \cite{caldi, utpal, linder, bhatt}
and they can also generate the Majorana mass term. 
In this work we investigate this scenario further with the aim of 
investigating  experimental signatures of
the background neutrino-condensates. 

Our arguments are based on the phenomenon called the "Andreev reflection" 
which is extensively studied
in the literatures of superconductivity \cite{andreev}.
 Before explaining the Andreev reflection,
we first mention the general ideas for the neutrino condensation.
Our analysis is based on a very simple and
elegant model given by Caldi and Chodos \cite{caldi}, although the 
results are more general and 
can be applied to the other mechanisms also. This picture requires 
that the background neutrinos 
to be in a degenerate gas/liquid state. The idea of degenerate 
neutrinos in the universe is not inconsistent
with the standard model of cosmology\cite{wein,dolgov}. It is well 
known that at a low enough
temperature a gas of degenerate fermions undergo a 
phase-transition to the
superfluid state, if there is an attractive-interaction between 
the fermions in
any angular-momentum channel\cite{juerg}. However such
 transition for the Dirac neutrinos
within the standard electroweak interaction may not be 
possible \cite{caldi}(see also Ref.\cite{kapusta}).
Interestingly, they also suggest that the degenerate 
Dirac-neutrinos in the presence of
some new (unknown) attractive interaction, howsoever weak, can
form condensates. 

The condensation can be studied by using the Lagrangian 
density for the
four-fermion interaction within  the the mean-field 
theory frame work \cite{caldi,utpal, kapusta}. 
From this one can find the BCS-type superfluid condensation 
with the gap parameter given by
$\Delta \sim \mu e^{-\frac{1}{\mu^2G^2}}$,
where $\mu$ is the neutrino chemical potential and $G$ signifies the 
strength of the four-fermions coupling.
Such condensates can naturally give rise to Majorana
mass of the neutrinos. In fact for the values of $\mu$ allowed
by the big-bang nucleosynthesis, it is
always possible to choose the values
of $G$ such that the cosmological constant, 
$\Lambda\sim~G^2|<\nu\nu>|^2$, is of the
order $(10^{-3} eV)^4=(meV)^4$ and the value of neutrino mass,
$m_{\nu}\sim\Delta~=~G^2|<\nu\nu>|$, can be anywhere between
a $meV$ and few hundred $MeV$.
Neutrino mass generated through such a mechanism will not
have any significant contribution to the dark matter in the universe.
However, if we allow the chemical potentials for different neutrino
species to vary, the condensates could depend non-trivially on
flavor, leading to an interesting spectrum of neutrino masses
and mixing.
This suggests a  possible mechanism for neutrino oscillations without
violation of the lepton number conservation at the microscopic level.

\section{Andreev Reflection}
It is not difficult to understand the Andreev reflection
phenomenon qualitatively.
Generally the quasi-particle
states propagating in a superfluid with energy
$\epsilon$ below the gap energy $\Delta$
are forbidden. But the sub-$\Delta$ transfer is
possible if an incoming electron with $\epsilon$
is transfered along with an another electron by
forming a Cooper pair into the superconductor.
In terms of single particle state this can be described
by the reflection of a hole in the normal metal.
Thus the Andreev reflection of a hole (or an electron)
is equivalent to the transfer of a
Cooper pair in (or out of) the superconductor. But the
single electron transfer with the sub-gap energy
is not allowed in the superconductor. Quasi-particles
of energy $\epsilon>\Delta$, propagate in the condensate
as a massive particles, of mass equal to $\Delta$
\cite{andreev}.\\
In the mean field description, the starting point is
an effective Hamiltonian with four-fermion interaction given by,
\begin{eqnarray}
H=\int d^3x\Big[\sum_{\alpha=\uparrow,\downarrow}
\psi^{\dagger}_{\alpha}(r,t)\Big(-\frac{\nabla^2}{2m}-E_F\Big)
\psi_{\alpha}(r,t)\nonumber\\
+\Delta(r,t)\psi^{\dagger}_{\uparrow}(r,t)\psi^{\dagger}_{\downarrow}(r,t)
+\Delta^{\star}(r,t)\psi_{\downarrow}(r,t)\psi_{\uparrow}(r,t)\Big]
\end{eqnarray}
where, $E_F$ is the Fermi
energy ($\sim$ chemical potential), and the "order parameter"
field $\Delta$ is
\begin{eqnarray}
\Delta_{\alpha \beta}(r,t)=<\psi_{\alpha}(r,t)\psi_{\beta}(r,t)>
=\Delta(r,t)\epsilon_{\alpha \beta}
\end{eqnarray}
The equation of motion obtained from the Hamiltonian above is known as
Bogoliubov- de Gennes equation in the theory of superconductivity/
superfluidity and is given by,
\begin{eqnarray}
\imath\partial_t \left( \begin{array}{ccc}
\psi_{\uparrow}(r,t)\\
\psi_{\downarrow}^{\dagger}(r,t)\\
\end{array}\right)
=\left( \begin{array}{ccc}
-\frac{\nabla^2}{2m}-E_F & \Delta(r,t)\\
\Delta^{\star}(r,t) &\frac{\nabla^2}{2m}+E_F\\
\end{array}\right)
\left( \begin{array}{ccc}
\psi_{\uparrow}(r,t)\\
\psi_{\downarrow}^{\dagger}(r,t)\\
\end{array}\right)      \nonumber\\
\end{eqnarray}
These linear equations can be treated
as equations for single particle wave functions instead of operators.
For $\Delta = 0$, the equations decouple and we have the plane wave
solutions,
with the dispersion relation, $E=\epsilon_q$ for particles and
$E=-\epsilon_q$ for holes (antiparticles),
$\epsilon_q=\vec{q}~^2/2m-E_F$. This is the case with the normal
conductors (metals).
When $\Delta=const.\neq 0$, we still have plane waves as solutions,
however, with the change that
quasiparticle spectrum within the condensate is gapped,
\begin{eqnarray}
E^2=\epsilon_{q}^{2}+|\Delta|^2
\end{eqnarray}
The quasiparticle wave functions are,
\begin{eqnarray}
\left( \begin{array}{ccc}
\phi_{p}\\
\phi_{h}\\
\end{array}\right)=
&&D\left( \begin{array}{ccc}
A_{+}~exp(\imath\delta)\\
A_{-}~exp(-\imath\delta)\\
\end{array}\right)exp(\imath q_{+}\cdot r-\imath E t)+\nonumber\\
&&F\left( \begin{array}{ccc}
A_{-}~exp(\imath\delta)\\
A_{+}~exp(-\imath\delta)\\
\end{array}\right)exp(\imath q_{-}\cdot r-\imath E t)
\end{eqnarray}
where $A_{\pm}=\sqrt{\frac{1}{2}(1 \pm \xi/E)}$,
$\xi=\sqrt{E^2-|\Delta|^2}$,
$q_{\pm}^2/2m=E_{F}\pm \xi$, $\delta$ is the phase of the complex order
parameter, and $D$ and $F$ are some constants. The wave functions,
$\phi_{p}$ and $\phi_{h}$ correspond to particle and hole like excitations
of mass $\Delta$.
However, note that they are not in the eigenstate of particle number
or electric charge.
For $E>>|\Delta|$, they reduce to wave functions for particles
and holes respectively. When the quasi-particle excitation energy,
$E<|\Delta|$, the momenta are complex, and therefore, these modes
do not propagate in the condensate
medium.
When such excitations reach the normal conductor-condensate junction
from the conductor
side of the arrangement, we have, what is known as Andreev reflection
in the theory of superconductivity.\\
Let us consider a normal conductor-superconductor
slab aligned along the $z-axis$.
Assume $\vec{r}=(0,0,z)$.
The junction of the materials is located at $z=0$.
At the conductor side ($z<0$), $\Delta=0$  while at the
superconductor side ($z>0$), $\Delta=constant$.
Let us now suppose that particles of energy, $E$, from conductor side
of the slab, hits the junction.
To see the details of what happens  at the boundary,
we need to solve
the stationary Bogoliubov - de Gennes equation, Eq.(3),
with the boundary conditions
: (1) for $z\rightarrow -\infty$ excitations are particles
or holes (conductors), (2) for $z\rightarrow \infty$ excitations are
quasi-particles (superconductor), (3) for $z=0$
wave functions and their first spatial derivatives are continuous at
the boundary at $z=0$. It is not hard to solve this
problem and detailed solutions can be found in reference \cite{andreev}.
The net result is that the probability current through the junction is,
\begin{eqnarray}
j_{z}=\left\{ \begin{array}{ccc}
0 & \mbox{for $E<\Delta$}\\
\frac{2\xi}{E+\xi}+\bigcirc(\frac{1}{E_F}) & \mbox{for $E>\Delta$}\\
\end{array}\right.
\end{eqnarray}
where $v_F=\sqrt{2E_F/m}$, is the Fermi velocity. The
reflection and transmission coefficients are,
\begin{eqnarray}
\left\{ \begin{array}{ccc}
R_{hole}=1, & T_{quasi}=0 & \mbox{for $E<\Delta$}\\
R_{hole}=\frac{E-\xi}{E+\xi}, & T_{quasi}=\frac{2\xi}{E+\xi} &
\mbox{for $E>\Delta$}\\
\end{array}\right.
\end{eqnarray}
These equations clearly show that an incoming particle hitting the junction
from the conductor side with energy below the gap can not cross the junction,
instead a a hole is reflected back in the opposite direction. This is not
difficult to understand physically- near the boundary, the incoming particle
creates a particle and a hole of equal and opposite momenta from the Fermi sea,
pairs up with the particle and disappears in the condensate and the hole is
reflected back. This is the Andreev reflection phenomenon. For $E>>\Delta$,
$R_{hole}\rightarrow 0, T_{quasi}\rightarrow 1$. For energy, $E$, not too high
above the gap, $\Delta$, there
is partial transmission of quasiparticles and partial reflection of holes.\\
\section{Experimental Signatures}
To understand the role of Andreev reflection in the context of neutrinos,
we have to find the fate of the condensate within materials.
%
%
Cosmological neutrinos are of low energy and large wavelength, and
threfore, the size of the cooper pairs in the condensate would be
large.
Weak interaction of neutrinos with matter is mediated through charge currents
as well as neutral currents.
However, the cross section of such interactions is
extremely small.
For $E_{\nu}=1~ MeV$, this cross section is $\sim 10^{-44} cm^2$, and for
$E_{\nu}=1~ eV$, it is $\sim 10^{-56} cm^2$.
Thus it would seem that matter is almost transparent for neutrinos.
However, it turns out that for low energy neutrinos there is coherent
neutrino processes in matter\cite{lewis}.
For momentum $\sim$ a few
hundred $MeV/c$, neutrinos would elastically scatter from nuclei with
cross section proportional to $A^2$. For lower momenta it will coherently
scatter off the entire atom, meaning all the nucleons and electrons in the
atom. Subsequent lowering of momenta involves, coherent scattering of
neutrinos from all the atoms contained in the wavelength. Thus the scattering
cross section would grow as the square of number elementary particles
( quarks $\&$ leptons) contained within the wavelength. Thus, it would seem
that the cooper pairs could be broken and condensate could be disordered
by coherent scattering processes.
We know from work of Caldi and Chodos \cite{caldi}
that the value of the order parameter is decided by the chemical
potential and the coupling constant of the assumed "new interaction".
From the big bang nucleosynthesis arguments, we get only the upper bound
on the combined value of the chemical potentials of all the three flavors
of neutrinos. We may get an estimate of the range of the "new coupling" from
the formula for the neutrino mass and the cosmological constant
obtained from condensate order parameter.
The neutrino mass is given by, $m_{\nu}\sim \Delta=G^2 |<\nu\nu>|$ and
cosmological constant by, $\Lambda\sim~G^2|<\nu\nu>|^2$, and therefore,
$G^2 \sim m_{\nu}^2/\Lambda$. If $m_{\nu}\sim meV$
($\sim 10^{-3}~eV$) and
$\Lambda\sim (meV)^4$, range of the coupling $G$, is of the order of $meV$.
But for $m_{\nu}\sim eV$, the range changes to micro electron volts.
Moreover, for different flavor of neutrinos, we expect that the
chemical potentials as well as the coupling constants are slightly different.
Thus there is large uncertinity in the values of the relevant parameters
and it is hard to decide whether the cooper pairs are stable or disordered
by coherent scattering processes within ordinary materials.
Therefore, for experimental signatures of condensates,
we will consider both the possibilities.
In the first case, the cooper pairs are assumed to be broken
within ordinary materials and the
material surface forms boundary with the cosmological condensate.
In the second case, we assume that the condensate
penetrates ordinary matter and the cooper pairs are intact.\\
We have, so far, considered only the neutrinos of cosmological
origin. However, there exist many other sources of neutrinos
in nature as well laboratory.
The energy spectrum of
neutrinos produced in these processes is very wide,
but most of the flux is concentrated in
energy of the order of a few MeV.
With the cosmological context in mind,
we are interested in properties of low energy neutrinos.
It turns out that the
radioactive decay of tritium contains low energy flux of antineutrinos.
We will, therefore, consider an experimental set up with
radioactive tritium sample undergoing beta decay. The experimental set up
should be conceptually similar to the arrangement used in the
KATRIN experiment \cite{katrin}.
The tritium nucleus decays to,
$^3H_1\rightarrow~^{3}He_{2}+e^{-}+\bar{\nu_e}$, with half life of
$T_{1/2}\approx 12.3$ years. The total energy released in the decay
is $E_0=18.6~ KeV$. Beta decay energy spectrum is analyzed
by using the Kurie plot \cite{giunt}.
The electron spectrum in the allowed beta decay
is,
\begin{eqnarray}
N_{e}(E_e)dE_e &\propto& F(Z,E_e)\sqrt{E_{e}^2-m_{e}^2}E_e(E_0-E_e)\nonumber\\
&\times& \sqrt{(E_0-E_e)^2-m_{\nu}^2}~dE_e
\end{eqnarray}
where $F(Z,E_e)$ is the known Coulomb factor, $E_0$ is the total
energy released
in the beta decay of the nucleus,
$E_e$, is the energy carried away by the electrons.
The plot of the Kurie function, $[N_{e}(E_e)/(F(Z,E_e)p_eE_e)]^{1/2}$
versus electron energy $E_e$ should be a straight line when $m_{\nu}=0$
but should be of a different shape near the end point when $m_{\nu}\neq 0$.
We are interested in the properties of the neutrinos released in the beta deacy
assuming the presence of cosmological background neutrino condensation.
For clarity and simplicity,
we assume that the background
cosmological condensate is made from cooper pairs of antineutrinos.
For neutrino condensates, one can use essentially the same arguments
in a slightly different manner.
In this situation, we have the following two possiblities:\\

\noindent
(1). The cooper pairs of the cosmological condensate
are formed by antineutrinos
of low energy and large wavelength. The pair binding is not
too strong even by the standard of weak interactions, and
therefore, within the radioactive material (tritium) the
condensate is disordered by the process of coherent scattering.
The condensate outside the material remains unchanged.
We have antineutrinos both from the disordered
cosmological condensate and the
the radioactive beta decay within the tritium sample.
We encounter here a situation similar to one in condensed matter system
with metal-condensate boundary. We, therefore, refer to
Andreev reflection phenomenon to understand the physical processes
in the material, condensate and their boundary.
Antineutrinos ( and neutrinos if any)
of energy lower than the value of the condensate order parameter $|\Delta|$,
are trapped inside the material. If any of these sub-barrier antineutrinos,
of momentum $\vec{p}$,
reaches the condensate boundary, a neutrino-antineutrino pair
of equal and opposite momentum, $\vec{-p}$ and $\vec{p}$,
is created at the junction. The newly created
antineutrino combines with the old incoming antineutrino,
forms a cooper pair and disappears in the condensate.
The neutrino (hole) is reflected back. Similar phenomenon takes place
when the reflected neutrino reaches the boundary.
However, those with energy greater than $|\Delta|$, move into
the condensate and propagate as quasiparticles
of mass $|\Delta|$ , their dispersion relation being,
$E=\sqrt{p^2+|\Delta|^2}$.
Thus the energy of the antineutrinos from the beta decay in the
sample covers
the entire spectrum of the released energy
and thus the Kurie plot in this case
is a straight line (as in the case of massless neutrinos)
with end point at the maximum energy, $E_0=18.6 KeV$
(with a few extra events near the end point which
will be explained latter).
Such a mechanism of mass generation does not lead to violation of
lepton number at the microscopic level,
and therefore, also implies that there
is no neutrinoless double beta decay \cite{doublebeta}.\\
Andreev reflection process creates a mixture of
sub-barrier energy neutrinos (holes )
and antineutrinos within the sample. The process
$^3H_1+\nu\rightarrow$ $^3He_2+e^{-}$
is energetically feasible and has the same cross-section as
$^3H_1\rightarrow$ $^3He_2+e^{-}+\bar{\nu}$.
Reaction cross section for such a processes
involving low energy neutrinos is very small.
However, in an experiment run over very long time
some low energy antineutrino excess event, which would show
up as excess electrons near the end point of Kurie plot,
should be seen. The estimated number of such electrons is
approximately $50\%$ more than what is expected from the standard
model.
From the end point of the spectrum where the excess events start
appearing, one can estimate the value of the order parameter.
It should be noted that the background cosmological
neutrinos, in the
absence of condensation would also create excess events near the end of
of the beta deacy spectrum, however unlike in our case, the energy spectrum
will have a monoenergetic peak at $E=E_0+m_{\nu}$, where
$E_0$ is end point of spectrum and $m_{\nu}$ is the neutrino mass \cite{kaboth}.
Experiments such as KATRIN is designed
to search for beta decay to low energy antineutrinos and it is
in these experiments that one would expect such events to be
established or ruled out.\\

\noindent
(2). The cosmological condensate penetrates the tritium sample and
there is no change either in the superfluid ordering or the value of the
order parameter.
As explained earlier, single particle
states of energy less than the value of the order parameter
can not be sustained in the
condensate. Because of this constraint on the available phase space,
radioactive
tritium can not decay into sub-barrier energy antineutrinos. However,
it can decay into antineutrinos of energy
greater than the value of the order parameter, $|\Delta|$.
These supra-barrier neutrinos produced in the beta decay
in the nucleus have wavelength shorter than
the neutrinos in the cooper pair
and they are
in eigenstate state of particle number operator within the cooper pair length scale.
Therefore,
in the calculation of beta decay amplitude they are considered massless,
with dispersion relation,
$E_{\nu}=cp_{\nu}$ (elsewhere, $c=1$).
In the derivation of Kurie function in this case, the important point
that we have to keep in mind is that $E_{\nu}>\Delta$, because
there is no single particle state below this energy.
Thus the Kurie plot is still a straight-line but the end point has a
cut off at energy equal to $E_0-|\Delta|$.
However, within the condensate these antineutrinos will propagate
as quasi-particles (in Bogoluobov state)
of mass $|\Delta|$, with  dispersion relation given by,
$E=\sqrt{p^2+|\Delta|^2}$. As in the previous case, the
the antineutrinos (and neutrinos) propagate as
massive quasi-particles having  Majorana mass but
there is no violation of the lepton number conservation at the
microscopic level, and therefore, there is no possibility of
neutrinoless double beta decay.
The massive quasi-particles are not in the eigenstate states of particle number
or the electronic lepton number operators. This is also
the necessary requirement for
neutrino oscillations.Thus, slightly deferent values
of the condensate order parameters for three flavors of neutrinos
would nicely account for neutrino oscillations.\\
In both the cases discussed above, we find that the value of the order
parameter (and thus the mass of the neutrinos) can be estimated from the
end point behaviour  of the beta decay spectrum. The rate of neutrinoless
double beta decay is known to be proportional to the square of neutrino
mass. Thus, we can estimate this decay rate and expect that in a suitably
designed experiment, it can either be established or ruled out.
%
%
\section{Conclusions}
We have discussed in some details how the superfluid
condensation of background neutrinos
of cosmological origin can generate neutrino mass for
flavor oscillations and also
account for the accelarated expansion of the universe.
Such a mechanism does not seem to be
in conflict with any aspect of the standard model of particle
physics. However, it indicates the existence of new physics in the
form of new attractive interaction among the neutrinos. We have shown
that, in carefully prepared beta decay experiments,
this model would lead to experimental signatures as listed below:\\
\noindent
(i). Absence of neutrinoless double beta decay is a necessary requirement.
If neutrinoless double beta decay is observed, the mechanism of mass
generation and oscillation discussed in this paper should be considered
redundant.\\
\noindent
(ii). The Kurie plot is a straight line with either of the two possibilities,
(a) the end point of the plot extends up to the total beta decay energy
released, $E_0=18.6 KeV$, with a few extra low energy events in an
experiment run over sufficiently long time (such events are known to exist,
however, they are still within the experimental error bars and can
be settled only by future experiments \cite{giunt}),
(b) the straight line terminates at $E_0-|\Delta|$, where $|\Delta|$ is the
superfluid order parameter.\\
Thus, it should be possible to either establish
or rule out cosmological neutrino condensation from carefully
prepared beta decay experiments in laboratory.

%

\end{document}